\documentclass[oribibl]{llncs}
\pdfoutput=1
\usepackage{paper}

\title{Simulating Ability\rlap{:}\\%
Representing Skills in Games}
\titlerunning{Simulating Ability}

\author{Magnus Lie Hetland}
\authorrunning{M. L. Hetland}

\institute{Norwegian University of Science and Technology,\\
    Trondheim, Norway\\
    \email{mlh@idi.ntnu.no}}

\begin{document}
\maketitle

\begin{abstract}
Throughout the history of games, representing the abilities of the various
agents acting on behalf of the players has been a central concern. With
increasingly sophisticated games emerging, these simulations have become more
realistic, but the underlying mechanisms are still, to a large extent, of an
ad hoc nature. This paper proposes using a logistic model from psychometrics
as a unified mechanism for task resolution in simulation-oriented games.
\end{abstract}


\section{Introduction}

One of the most fundamental concept in games is the representation of agents,
entities acting on behalf of a player, and the simulation of their abilities.
Consider, for example, the different pieces of chess, and the moves they are
capable of; or the troops in the game of Risk~\citep{Risk:1959}, and their
offensive and defensive strengths, represented by the number of dice to roll.
In earlier games such as these, the game mechanics tend to treat abilities in
a rather abstract and simplified manner, but as games have become more true to
life, so has the simulation of skills. One important line of development in
this area started with the tabletop roleplaying games, such as the original
Dungeons \& Dragons~\citep{Gygax:1974}, and has continued to the present day,
with computer games requiring skill systems of high sophistication. Yet little
has been published about such skill systems, or what constitutes realistic
simulation of skills and abilities. This paper is an attempt at drawing
parallels between skill simulation in games, on the one hand, and skill
modeling in psychometrics on the other, arguing that a simple log-odds
model is an attractive alternative to many of the ad hoc systems that have
been used so far.

\section{Skills and Abilities in Games}

In this section, I briefly discuss some existing approaches to skill modeling.
I start by examining a variety of models that have been used in tabletop
roleplaying games. This is motivated in part by the fact that descriptions of
these models are much more readily available than the ones embedded in
proprietary game code, and in part by the fact that ``Most [computer
roleplaying games] use a system based on an old paper [roleplaying game] to
handle their game mechanic''~\citep[p.~358]{Rollings:2003}. After outlining
the main models, I describe some flaws in the most commonly used one, and
outline an alternative.

\subsection{Some Classic Approaches}

A survey of published roleplaying games indicates that a few mechanisms are by
far the most common. There is signiticant variation in the details, but even
so, a few main models may be distilled.%
\pagenote{This survey was based on the encyclopedic listing of published
roleplaying games (currently \si{1524} games) maintained by
\citet{Kim:2012}. As the focus of the survey was die mechanics, games
using cards, tables, pebbles, roulette wheels or other randomness
generators, as well as diceless games (that is, ones without randomness)
were eliminated. Certain die mechanics that strayed too far from the
independent simulation of a single action (for example, ones including
dice that are spent and recuperated over time, or games that mix the use
of dice with various other narrative mechanics) were also removed. This
initial elimination accounted for \si{129} games (or about
\SI{8}{\percent}). Games whose die mechanic was not unambiguously
described or readily available were also eliminated (another \si{652}
games, or \SI{42}{\percent}). There is no clear reason to assume that the
elimination of these latter games represents any strong bias, and although
the criteria used required some amount of subjective judgement, it is
assumed that the remaining \si{729} games (\SI{48}{\percent}) form a
sample that is representative enough for the current purposes.}

\paragraph{Uniform scale.}
The most popular mechanic is the uniform scale. While it has different
implementations in terms of die rolls, the underlying model is this: Skill and
difficulty are both represented on the same scale, and the probability of
success is a linear function of their difference. One common implementation is
treating the skill as a fixed probability (or a fixed value on a uniformly
random scale), and the difficulty as an additive modifier. A uniformly random
value is then generated using dice, and if this value is equal to or less than
the modified skill, the action is a success.
Another common implementation is to treat the skill as an additive modifier to
the uniform die roll, trying to roll a sum that exceeds a difficulty level.
Various variations are of course possible here, but they amount to the same
thing, that is, a linear mapping from the difference $\mathit{skill} -
\mathit{difficulty}$ to the probability of success, capped at \si{0} and
\SI{100}{\percent}. Alternatively, it can be viewed as simulating the outcome
as being uniformly distributed, usually symmetrically around the skill level.
The difficulty is then the outcome quality required to succeed.\footnote{Note
that this is not necessarily how outcome quality is simulated in these games.}

\paragraph{Sum of dice.} This simply means that you roll a fixed number of
identical dice to generate your distribution. The uniform scale is the special
case you get with a single die; with even two dice, the distribution becomes
decidedly more bell-curved. This mechanism is used in about one in five games,
with both the number and type of dice varying quite a bit. The two main
implementations discussed for the uniform scale (roll under skill or roll over
difficulty) are both used here as well.

\paragraph{Binomial die pool.} The simplest version of this is flipping a
number of coins equal to the skill level, and counting the number of heads.
More generally, a number of dice representing the skill level are rolled,
and each die exceeding a given threshold value counts as a success, usually
with multiple successes representing a better outcome. In some
games, difficulty is represented by the target number, while in others, the
target is fixed, and difficulty is represented as the required number of
successes.

\paragraph{General die pool.} This is a broader class of systems that includes
games where the ability is represented as a number of dice, and the outcome is
the sum of those dice. This is, of course, a generalization of the binomial
case, except that the distribution for each die is fixed (as opposed to the
Bernoulli trials in the binomial case).

\paragraph{Step dice.} This is similar to the uniform case, as the outcome
variable is, indeed, uniform. However, instead of modeling skill as an
additive offset, it is treated as multiplicative; or, rather, the skill is
mapped to the type of die, which dictates the range of outcomes.

\paragraph{Max-die pool.} This is similar to the die pool mechanic, but rather
than adding the dice, the maximum is used. That is, the skill indicates the
number of dice to roll, and the highest die roll indicates the outcome.

\bigskip\noindent
There are, of course, other mechanics, but most games fall into one of these
categories, and of them, the uniform scale is clearly the most commonly used,
accounting for about half the games surveyed.

\subsection{The Problem With Percentiles}\label{sec:percentiles}

The uniform distribution certainly has an intuitive appeal. A superficial
analysis seems to indicate that, as opposed to for the bell curve, a fixed
modifier such as $+\SI{10}{\percent}$ means the same whether your starting
probability (your skill) is high or low. This analysis quickly breaks down
when looking at the edge cases. A bonus of $+\SI{10}{\percent}$ is clearly
more useful to anyone with a skill of $\SI{90}{\percent}$ or less than to
those more skilled.

We might, instead, consider an \emph{unlimited} scale, where $+10$ really
\emph{does} always mean $+10$, for example.\footnote{Exactly what this means
will be explained in the following.} If we ignore randomness, and assume that
skill level is all that matters for the outcome, a skill level of $a$ will
always beat a skill level (or difficulty level) of $b$, if $a > b$. This is in
a sense the assumption underlying so-called Guttman scales, and the model
discussed in the following sections is an extension to this, to account for
random variations of various kinds.

\begin{figure}
\def\thecircle{(axis cs:-37.5,137.5) circle (142.52192813739262)}
\begin{center}
\begin{tikzpicture}
    \begin{axis}[
        width=7cm, height=7cm,
        xmin=0, ymin=0, xmax=100, ymax=100,
        xlabel={Easy task},
        ylabel={Hard task},
        xtick={0,25,50,75,100},
        ytick={0,25,50,75,100},
        x unit={\si{\percent}},
        y unit={\si{\percent}},
    ]
        \addplot[dashed] coordinates {
            (25,0)
            (100,75)
        };
        \draw[gray]
            (axis cs:50,0) --
            (axis cs:50,25) --
            (axis cs:0,25)
            ;
        \draw[gray]
            (axis cs:75,0) --
            (axis cs:75,50) --
            (axis cs:0,50)
            ;
        \draw \thecircle;
        \draw (axis cs:0,51) node[above right,font=\footnotesize] {Person A};
        \draw (axis cs:0,26) node[above right,font=\footnotesize] {Person B};
    \end{axis}
\end{tikzpicture}
\caption{Basic constraints on how task difficulty must work necessarily lead
to a non-linear probability scale}\label{fig:equating}
\end{center}
\end{figure}
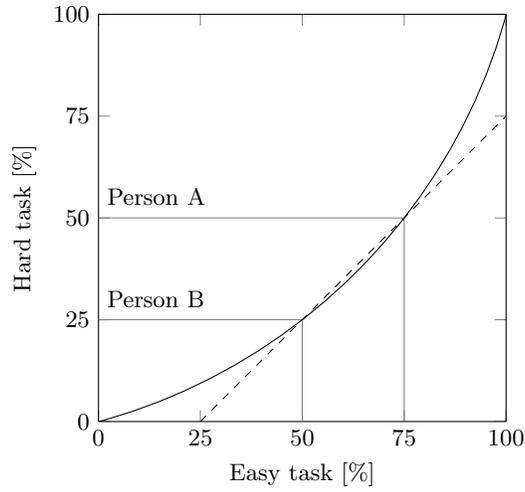

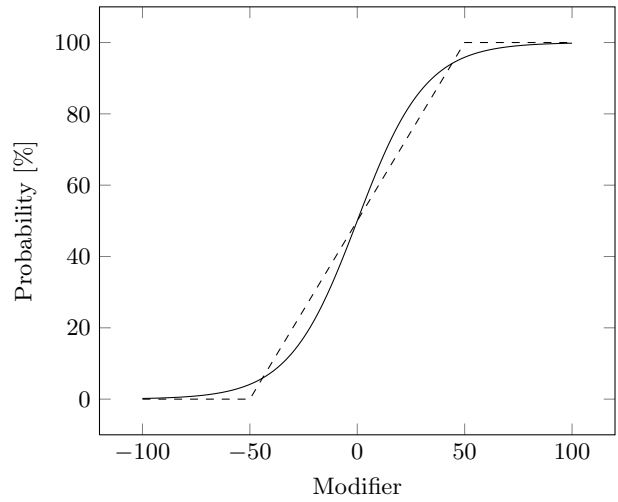
\begin{figure}
\begin{center}
\begin{tikzpicture}
    \begin{axis}[
        xlabel=Modifier, ylabel=Probability,
        y unit={\si{\percent}}
    ]
        \addplot[black,mark=none] table {logistic_vs_uniform.txt};
        \addplot[dashed,mark=none] coordinates {
            (-100,0)
            (-50,0)
            (50,100)
            (100,100)
        };
    \end{axis}
\end{tikzpicture}
\end{center}
\caption{The cumulative uniform distribution as an approximation to the
cumulative logistic distribution (with matched
variance), for modifiers to a skill level with an initial probability of
success of \SI{50}{\percent}}\label{fig:piecewiselinear}
\end{figure}

\Citeauthor{Wright:1993} highlights the problems with the uniform scale (in
the context of test equating) as follows~\citep{Wright:1993}.
Consider two tasks of the same kind (involving the same skill), one easy
and one hard. Let's say two persons, A and B, attempt both tasks. In general,
we'd expect them to succeed more often on the easy task than on the hard task,
except if they're at the extremes (succeeding \SI{0}{\percent} of the time on
the easy task, or \SI{100}{\percent} on the hard one). We could also assume
that A is more skilled B, and should therefore succeed more often than B
on both tasks. As shown in Fig.~\ref{fig:equating}, our assumptions lead to the
need for a non-linear mapping between the two; the uniform, linear model gives
us some problematic thresholding effects.
Figure~\ref{fig:piecewiselinear} illustrates this point by comparing the piecewise
linear curve resulting from the uniform model with a smooth, logistic curve.
The latter, I will argue in the following, is a much better choice for skill
modeling in general.

\subsection{The Outline of a Model}

The following brief line of reasoning is based on \citeauthor{Rasch:1980}'s
motivation for his eponymous probabilistic model of skill levels and
difficulty~\citep[pp.~72--75]{Rasch:1980}. While it does not cover all the
details we'll be investigating, it does give us an outline of the form our
skill model should take.\pagenote{A more thorough discussion of this topic,
with the requirements of so-called invariant measurement spelled out in
detail, can be found in the recent book on this topic by
\citet[p.~13--17]{Engelhard:2012}.}

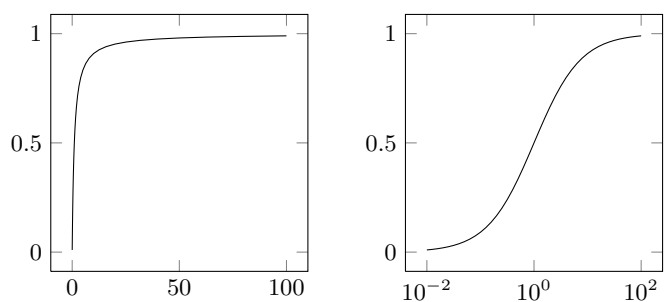
\begin{figure}
\begin{center}
\begin{tikzpicture}[baseline]
    \begin{axis}[
            width=5cm, height=5cm,
        ]
        \addplot[black,mark=none] table {odds_plot.txt};
    \end{axis}
\end{tikzpicture}
\hspace{0.15cm}
\begin{tikzpicture}[baseline]
    \begin{semilogxaxis}[
            width=5cm, height=5cm,
        ]
        \addplot[black,mark=none] table {odds_plot.txt};
    \end{semilogxaxis}
\end{tikzpicture}
\caption{The probability of success as a function of the skill-to-difficulty
ratio%
}\label{fig:odds}
\end{center}
\end{figure}

What we're seeking is a function $f$ that takes a skill level $a$ and a
difficulty level $x$ and produces a probability of success, $f(a, x)$. If we
assume that both $a$ and $x$ are measured on a multiplicative scale, we'd
expect a person that is twice as skilled to be able to deal with tasks that
are twice as hard, or, in general, that $f(a,x) = f(ka,kx)$, for any positive
constant $k$. Now, consider person A and person B, with respective skill
levels of $a$ and $b$, and problems X and Y with difficulty levels of $x$ and
$y$. Assume that A is $k$ times as skilled as $B$, that is,
$a = kb$,
and problem X is $k$ times as hard as problem Y,
$x = ky$.
This, of course, is equivalent to
$a/x = b/y$. In any such case, we'd want $f(a,x)=f(b,y)$. In other words, we
don't need to consider skill and difficulty separately, \emph{only their
ratio}, and we can write $f(a/x)$ rather than $f(a,x)$. For problems that are
way too hard (low ratio) we'd expect a probability near zero, and for problems
that are just too easy (high ratio), we'd want to get close to
a \SI{100}{\percent} chance of success. If, for simplicity (and without loss
of generality), we assumed that skill and difficulty are on the same scale, we
would also have $f(1)=\SI{50}{\percent}$. If skill and difficulty are to be
interchangeable (as, for example, when the skills of two persons are pitted
against each other), we would also require the probability to be symmetric,
that is,
\[
    f(a/x) = 1 - f(x/a)\,.
\]
Even given these desired properties, we have some leeway in choosing the exact
form of $f$. One function satisfying the requirements is shown in
Fig.~\ref{fig:odds}. The details of, and motivation for, this specific
function are discussed in the following.

\todo{More on how we might have some leeway, but that we basically need a
sigmoid (on a logarithmic scale). Perhaps mention the specific semantics of
this particular function (interpreting $a/x$ as odds; twice as good wins twice
as often); forward reference to evidence/odds.}

\section{Skills as Evidence: the 1PL}\label{sec:evidence}

In this section, I show that a rather natural interpretation of skill and
difficulty as evidence for and against a positive outcome yields a model that
conforms to all the desired properties outlined so far.

\subsection{Bayes Factors and Odds Ratios}

We have seen that the probability of success should be computed as a function
of the skill-to-difficulty ratio $a/x$, and that this function must have some
basic properties that amount to getting a sigmoid curve when plotting it with
a logarithmic horizontal axis, as in Fig.~\ref{fig:odds}.
It would seem that we can choose whichever function we want, as long as it
looks sort of like Fig.~\ref{fig:odds}, and that the various functions may all fit
different situations with a varying degree of accuracy. If we want one
unifying model, however, there is a particular function that stands out:
\begin{equation}
    f(a, x) = f(a/x) = \frac{a/x}{1 + a/x} = \frac{a}{a + x}\,.
    \label{eq:oddsfunc}
\end{equation}
This is the function \citeauthor{Rasch:1980} used, because it was the simplest
function he knew of that had the desired properties
~\citep[p.~74]{Rasch:1980}.\footnote{That is, it was the simplest function he
knew of that increases from 0 to 1 as $a/x$ goes from 0 to $\infty$.}
Simplicity is not the only reason to favor this function, however. For one
thing, it admits of a very intuitive interpretation, namely that the odds of
success are \O{a}{x}. The same idea was used already by \citet{Zermelo:1929},
in modeling chess players. In his model, the odds of player A with skill level
$a$ winning over B, with skill $b$, were simply \O{a}{b}. So if player A was
ten times as skilled as B, she would also win ten times as often.
But beyond its simplicity and intuitive appeal, this model follows very
naturally from the interpretation of skill and difficulty as \emph{evidence},
that is, as facts that influence the probability of the outcome.

First, let's look at the core rule for combining new evidence with existing
knowledge: Bayes' theorem. When formulated in terms of odds, it can be written
as follows.
\[
    \mathrm{R} = \mathrm{PL}
\]
Here, $\mathrm{P}$ is the prior odds, $\mathrm{L}$ is the likelihood ratio
(or Bayes factor), and $\mathrm{R}$ is the revised odds. In other words,
$\mathrm{P}$ and $\mathrm{R}$ represent your degrees of belief (as odds) in a
given hypothesis before and after being presented by a given piece of
evidence, and $\mathrm{L}$ tells you how much more likely it would be to observe
that evidence if the hypothesis were true.

If we (for simplicity) assume even prior odds,\footnote{We could also
let the innate task difficulty be represented as the inital odds, for example;
it makes no difference to the calculations.} and a series of factors whose
strengths (likelihood ratios) are $\mathrm{L}_1\ldots\mathrm{L}_n$, the
revised odds are
\[
    \mathrm{R} = \mathrm{L}_1 \times \cdots \times \mathrm{L}_n\,.
\]
Any factor with strength $\mathrm{L}$ \emph{for} an outcome automatically has
the strength $1/\mathrm{L}$ \emph{against} that same outcome. So if we let
skill level and difficulty be the two only factors, acting for and against
success, respectively, we end  up with \citeauthor{Rasch:1980}'s formulation.
Similarly, if we let two people's skills be the only two factors, acting
against each other, we end up with \citeauthor{Zermelo:1929}'s formulation.

The previous arguments give
us a plausible interpretation of
\citeauthor{Rasch:1980}'s probability function for the skill-to-difficulty
ratio. But how much leeway do we have in our modeling here? If we wish to
interpret factors such as skill and difficulty as pieces of evidence, modeled
by their weight, whatever that might mean, how much freedom do we have in
choosing the specific mapping? As it turns out, not much. In fact,
\citet{Good:1968,Good:1985} makes a convincing case that this is the
only formalization possible. In fact, the model follows from the following
three requirements:%
\pagenote{The first requirement means that we can have a function $W(H\prov
E)$ for the weight of the evidence in favor of $H$ provided by $E$, and
that this would be a function of the likelihoods $P(E\given H)$ and
$P(E\given\neg H)$, that is,
\(
    W(H\prov E) = f(P(E\given H), P(E\given\neg H))
\),
for some function $f$. The second requirement means that, for some function
$g$, we can write
\(
    P(H\given E) = g(W(H\prov E), P(H))
\).
Combining the two, we get
\(
    P(H\given E) = g(f(P(E\given H), P(H\given\neg H)), P(H))
\).
For simplicitiy, we let $x=P(H)$, $y=P(E)$, and $z=P(H\given E)$, and get
\[
    z = g\biggl(f\biggl(y\cdot\frac{z}{x},\,
    y\cdot\frac{1-z}{1-x}\biggr),\, x\biggr)\,.
\]
From this, we see that $g$ is mathematically independent of $x$, and must
therefore depend on the value of $f$. Therefore, $f$ must be mathematically
independent of $y$, so it necessarily depends only on the ratio of its
arguments, that is, of the likelihood ratio $\mathrm{L} = P(E\given
H)/P(E\given\neg H)$. Also, $f$ must be strictly monotonic (obviously
increasing), because otherwise it would have the same value for two different
values of $z$, which would violate the previous equation. If we add the third
requirement (combination through multiplication), the odds ratio model
follows.}
\begin{enumerate}
    \item The weight of one piece of evidence should only depend on how likely
        the evidence is to have been present given success, and given failure.
    \item The odds of success should only depend on the prior odds and the
        combined weight of evidence for and against it.
    \item The weight of independent pieces of evidence is combined by
        multiplication.
\end{enumerate}
In \citeauthor{Good:1985}'s derivation, the combination in the third
requirement is actually by \emph{addition}, rather than multiplication, which
simply entails using a logarithmic transform, measuring everything in log-odds
units, or \emph{logits}. This also aligns quite well with the grades of
evidence described by \Citet[p.~432]{Jeffreys:1983}, as shown in
Table~\ref{tab:grades}. These grades are also based on the Bayes factor
$\mathrm{L}$, and follow a logarithmic progression.

The resulting probability function is the so-called \emph{logistic function},
which is shown compared with the uniform model in Fig.~\ref{fig:piecewiselinear},
and is what the right panel in Fig.~\ref{fig:odds} would have depicted, had the
horizontal axis been linear.

\begin{table*}
\newcommand{\lft}[1]{\rlap{$#1$}\phantom{10^{1.5}}}%
\caption{\Citeauthor{Jeffreys:1983}'s grades of evidence}\label{tab:grades}
\ra{1.2}%
\begin{tabularx}{\textwidth}{@{}llX@{}}\toprule%
Grade 0 \hspace{3mm} & $\phantom{10^{1.5} < \mathrm{L}}\llap{$\mathrm{L}$} < 1$ & The
evidence is \emph{against} the hypothesis.\\
Grade 1 & $\lft{1} < \mathrm{L} < 10^{0.5} $ & The evidence is barely worth a
mention.\\
Grade 2 & $10^{0.5} < \mathrm{L} < 10$ & The evidence is substantial.\\
Grade 3 & $\lft{10} < \mathrm{L} < 10^{1.5}$ \hspace{3mm} & The evidence is strong.\\
Grade 4 & $10^{1.5} < \mathrm{L} < 100$ & The evidence is very strong.\\
Grade 5 & $\lft{100} < \mathrm{L}$ & The evidence is decisive.\\
\bottomrule\end{tabularx}
\end{table*}

\subsection{IRT and Rasch Models}\label{sec:irtrasch}

The model we have derived by viewing skill and difficulty levels as strength
of evidence is, in fact, the one commonly known as the one-parameter logistic
model, or 1PL, in the field of psychometrics.\pagenote{There are some
philosophical differences between the Rasch model and the
1PL~\citep[p.~19]{Ayala:2009}, but they are mathematically equivalent, and
the philosophical differences apply when viewed in the context of
\emph{measurement}, as opposed to the current issue of simulation.} This model
was initially introduced by \citet{Rasch:1980}, but has since become a
foundation for both so-called invariant measurement in the social
sciences~\citep{Engelhard:2012} and in the item-response theory of
psychological testing~\citep{Ayala:2009}.

The requirements of invariant measurement constrain us to using a sigmoid
probability function on an additive scale, as discussed in the introduction.
It is possible to use, say, a probit scale, but the logit model is vastly more
common. While a very important reason for this is its ease of computation, as
we have seen in the preceding section, there are also philosophical and
methodological reasons for favoring it.

This use of a logistic distribution can be found elsewhere as well. For
example, the initial Elo model for rating chess players used a normal
distribution, and this is still used by FIDE, the World Chess Federation.
However, investigations by the United States Chess Federation have found that
a logistic curve is a better match to real-world outcomes, and so they have
switched to a logit-based Elo model.\footnote{See, e.g., ``Arpad Elo and the Elo
Rating System'' in the December 16, 2007 issue of \emph{Chess News}
$\langle$\url{http://en.chessbase.com/home/TabId/211/PostId/4004326}$\rangle$.}
There are other player skill estimation systems, such as Microsoft's
TrueSkill ranking and matching
system,\footnote{$\langle$\url{http://research.microsoft.com/en-us/projects/trueskill}$\rangle$}
that use a normal distribution.

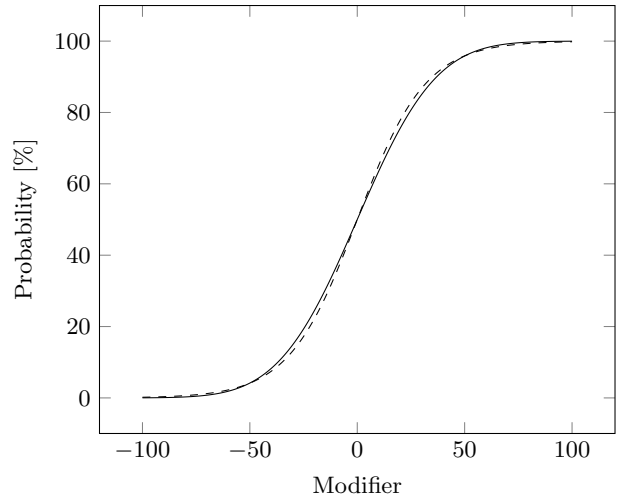
\begin{figure}
\begin{center}
\begin{tikzpicture}
    \begin{axis}[
        xlabel=Modifier, ylabel=Probability,
        y unit={\si{\percent}}
    ]
    \addplot[black,mark=none,dashed] table[y=logistic] {logistic_vs_normal.txt};
    \addplot[black,mark=none] table[y=phi] {logistic_vs_normal.txt};
    \end{axis}
\end{tikzpicture}
\end{center}
\caption{The cumulative normal distribution as an approximation to the
    cumulative logistic distribution (both with variances matched to the
    uniform distribution in Fig.~\ref{fig:piecewiselinear}), for modifiers to a
    skill level with an initial probability of success of
    \SI{50}{\percent}}\label{fig:normal}
\end{figure}

For our goals of approximate
simulation the difference may be of little practical importance though, given
that the greatest absolute difference between the two cumulative distributions
is less than \SI{1}{\percent}%
~\citep[p.~120]{Bowling:2009} (see Fig.~\ref{fig:normal}).

At this point it might be worth pointing out that even though the most
commonly used uniform model is a poor match for the kind of sigmoid we're
looking for, the second-most popular sum-of-dice mechanism is a rather good
fit. See, for example, Fig.~\ref{fig:dice}, which compares the distribution for
the sum of three six-sided dice (3d6) alongside a moment-matched logistic
curve.

\def\figdice{3d6\xspace}
\begin{figure}
\begin{center}
\begin{tikzpicture}
    \begin{axis}[
            xlabel=Outcome,
            ylabel=Cumulative probability,
            legend pos=south east,
            legend style={draw=none,font=\small},
        ]
        \addplot[
            jump mark left,
            black,mark=*,
            every mark/.append style={scale=0.7,fill=white},
        ] table {cum_dice_prob.txt};
        \addplot[black,dashed,mark=none] table {die_approx_logistic.txt};
        \legend{\figdice,logistic}
    \end{axis}
\end{tikzpicture}
\caption{The cumulative distribution of \figdice compared to a logistic curve
with matched mean and variance}\label{fig:dice}
\end{center}
\end{figure}
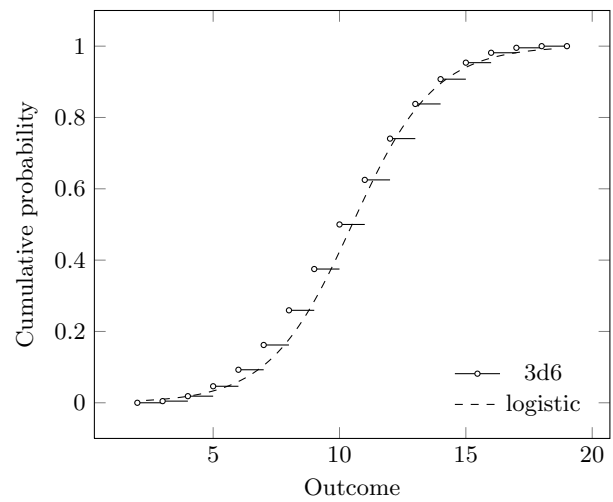
\todo{Mention continuity adjustment (+0.5).}

\section{The 2PL, 3PL and 4PL}

The 1PL nicely captures skill and difficulty as (additive) levels of evidence
for a successful outcome. There are aspects of task resolution that are still
not captured by this model, though, but that can be addressed by more flexible
psychometric models: the two-, three- and four-parameter models (2PL, 3PL and
4PL, respectively).

First, consider the choice of scale. We have chosen to use a single scale for
measuring skills and difficulties, but perhaps this is too restrictive?
Consider the following generalization of Bayes' theorem, described by
\citet{Zlotnick:1967}:
\[
    \mathrm{R} = \mathrm{PL}^r
\]
Here, $r\in[0,1]$ is a \emph{reliability rating} of the evidence represented by the
Bayes factor L. \Citeauthor{Zlotnick:1967} describes $r$ as the probability of
the reported evidence being real, as opposed to pure fabrication. For the
purpose of skill modeling, we could view $r$ as a degree of \emph{relevance}
between the modified skill level and the actual outcome. The lower the
relevance, the less impact the skill level will have, and the more random the
outcome. 
In our logit model, $r$ would simply be a factor, modifying the units
of the logit scale. If we relax the restriction $r\in[0,1]$, we end up with
the 2PL, where the scale or slope of the logistic curve is a separate
parameter.%
\footnote{In psychometrics, $r$ is commonly known as the \emph{discrimination
parameter}, describing the ability of a given task to discriminate between
high and low skill.}
Taking \citeauthor{Zermelo:1929}'s model of chess as an example, we
might want to apply an $r>1$ for the game of Go, and an $r<1$ for the game of
Ludo, representing the variability or degree of randomness inherent in the
games. For a game of ``heads-or-tails,'' we would set $r=0$, as the level of
gambling skill would be completely irrelevant, and the outcome is purely
random.

Some skill models (such as Microsoft's TrueSkill system) assume that the
variance in performance varies from person to person. In other words, the $r$
could be linked to the task at hand, to the person attempting it, or to the
relationship between the two (e.g., degree of relevance of the applied skill
to the task).

The 2PL is still a bit limited in that, unless we set $r=0$, the probability
of success for low and high modified skill still converges to \si{0}
and \SI{100}{\percent}, respectively. This is an issue that has been tackled
in psychometrics as well, in the context of guessing in multiple-choice tests.
The aim here is estimating the skill level of the students (in logits) based
on their test result, given that random guessing will give you the correct
answer to a question, say, \SI{25}{\percent} of the time. In this case, for
lower skill levels, the probability of success should converge on
\SI{25}{\percent}, not \SI{0}{\percent}. This requirement gives rise to the
three-parameter logistic model (3PL) of~\citet{Birnbaum:1968}. Generalizing to
the natural case where the upper asymptote may differ from 1 (that is, we can
have random failure), we end up with the four-parameter logistic model
(4PL)~\citep{Liao:2012}. This, then, is the model I propose for skill
simulation in games: that the probability of success for a skill level $a$ and
difficulty $x$ be given by
\begin{equation}
    f(a,x;r,\ell,u) = \ell + (u-\ell)\cdot\frac{1}{1+e^{-r(a-x)}}\,,
    \label{eq:4pl}
\end{equation}
where $a$ and $x$ are (additive) ability and difficulty, as before, $r$ is the
slope (indicating the reliability or relevance of the modified skill), and
$\ell$ and $u$ are the lower and upper asymptotes, respectively. In other
words, $\ell$ is the probability of failing randomly, regardless of skill,
while $1-u$ is the probability of succeeding randomly, regardless of skill.

\section{Discussion and Future Work}

In this paper I have argued that a four-parameter logistic model is a natural
choice when simulating skills in games. It matches simple intuitions of
bell-curve outcomes, it is backed up by viewing skills as evidence for
success, and it is a model used in psychometric practice, for estimating skill
levels based on successes and failures. Other, similar models are certainly
possible. For example, a Gaussian rather than logistic distribution could be
used, although the logistic distribution approximates the Gaussian very
closely, is much easier to work with, computationally, and leads to the
satisfying interpretation of skill levels and difficulty as weight of
evidence.

I have also argued for going beyond the basic Rasch model (1PL), parametrizing
the slope and asymptotes of the probability function. There is some
philosophical controversy about such additional parametrization in
psychometrics, related to the concept of invariant measurement, but these
don't really apply to the problem of skill simulation.

Future work on the model might include examining its suitability for modeling
outcome quality, rather than simply success and failure. This could be
relevant, for example, to modeling damage in combat simulations. To what
extent does it make sense to use the common die mechanic of simply generating
a value from the (logistic) probability distribution, adding it to the
modified skill, and viewing the result as an outcome quality? It might very
well be that the outcome measures don't map directly to the logistic scale;
this is, of course, an empirical question~\citep[see,
e.g.,][]{Reep:1971,Skinner:2009}.

One issue that has not been addressed in this work is the modeling of
\emph{player} skill, as opposed to \emph{character} skill. Models such as Elo
and TrueSkill are routinely used to match players of comparable skill levels.
The logistic model of character skill proposed in this paper is also used in
psychometric estimation. For games that employ both player and character
skill, a hybrid approach could be employed, where player skill is estimated,
and opposing characters are simulated with similar parameters, tuned to the
proper level of challenge.

Another important issue is skill development and learning. For games that
involve character development, it would be crucial to know how a skill,
measured in logits, would increase with repeated use and practice. It seems,
for example, that for many domains, the time to perform a task, as well as the
number of errors, follows a power law as a function of the number of
trials, the so-called ``power law of practice''~\citep{Newell:1980}. The
number of errors is, of course, directly linked to skill level in the logistic
model. This is also an area where player skill estimation could be useful, for
evaluating learning and skill gain over time.

\bigskip\noindent
\textbf{Acknowledgements.}\quad The author would like to thank Ole Edsberg for
fruitful discussions on the topic of this paper, including the role of
discernment and degree of randomness in task resolution, and for input on
existing player skill estimation systems.

\printnotes

\bibliography{paper}

\end{document}